\documentclass[aps,preprint]{revtex4}
\def\BibTeX{{\rm B\kern-.05em{\sc i\kern-.025em b}\kern-.T
    08em\kern-.1667em\lower.7ex\hbox{E}\kern-.125emX}}
\usepackage{graphicx}

\begin{document}
\title{Antimagnets: Controlling magnetic fields with superconductor-metamaterial hybrids}

\author{Alvaro Sanchez$^*$, Carles Navau, Jordi Prat, and Du-Xing Chen}
\affiliation{$^1$Grup d'Electromagnetisme, Departament de F\'{\i}sica, Universitat Aut\`onoma de Barcelona, 08193 Bellaterra,
Barcelona, Catalonia, Spain}


\begin{abstract}

{\bf

Magnetism is very important in science and technology, from magnetic recording to energy generation to trapping cold atoms. Physicists have managed to master magnetism - to create and manipulate magnetic fields- almost at will. Surprisingly, there is at least one property which until now has been elusive: how to 'switch off' the magnetic interaction of a magnetic material with existing magnetic fields without modifying them. Here we introduce the antimagnet, a design to conceal the magnetic response of a given volume from its exterior, without altering the external magnetic fields, somehow analogous to the recent theoretical proposals for cloaking electromagnetic waves with metamaterials. However, different from these devices requiring extreme material properties, our device is feasible and needs only two kinds of available materials: superconductors and isotropic magnetic materials. Antimagnets may have applications in magnetic-based medical techniques such as MRI or in reducing the magnetic signature of vessels or planes. 

}

\end{abstract}

\maketitle

\section{Introduction}
Cloaking a region in space from electromagnetic waves seemed something scientifically impossible until very recently. Pendry et al and Leonhardt \cite{controlling,leonhardt}, using the concepts of metamaterials and transformation optics, theoretically designed an electromagnetic cloak, which would render a given volume 'invisible' to electromagnetic waves. Such a device required extreme values of magnetic permeability $\mu$ and electrical permittivity $\varepsilon$. Although some experimental results presented  partial cloaking in special cases \cite{microwave,3D,natcomm}, no complete broadband cloak has been experimentally achieved until now \cite{leonhardtnature}.

In 2007 Wood and Pendry introduced the concept of magnetic cloaking \cite{wood}. They showed that in the dc case (electromagnetic waves in the limit of zero frequency), for which the electrical and magnetic effects decouple, a magnetic cloak for concealing static magnetic fields without disturbing the external field needed a material with anisotropic and position-dependent $\mu$ values, smaller than 1 in one direction and larger than 1 in the perpendicular one (see Fig. 1a and Eq. (1) for the case of a cylinder). A $\mu<1$ could be achieved by
arrays of superconducting plates  \cite{wood,ourAPL,magnus}, whereas $\mu>1$ could be obtained with ferromagnetic materials. However, no method has been presented to achieve in a real case the required position-dependent values in perpendicular directions simultaneously. Because of these difficulties a magnetic cloak has never been designed nor fabricated until now.

This eventual magnetic cloaking would have not only scientific interest but also important technological applications since magnetic fields are fundamental to many everyday technologies and in many of them it is necessary to have a precise spatial distribution of the magnetic field, which should not be perturbed by magnetic objects - not only by magnets but by any material containing iron or steel, for example. Is it therefore possible to build a cloak for static magnetic fields with - very important- using only ingredients that are practical and available? In this work we will demonstrate the affirmative answer to this question, by exploiting the properties of two worlds: metamaterials and superconductors.

\section{Definition of the antimagnet} 

At this point we want to redefine our goal into a more precise and even more ambitious one: instead of a magnetic cloak -null interior field and external field unaffected-, we want to design here an antimagnet, defined as a material forming a shell that encloses a given region in space while fulfilling the following two conditions:

i) The magnetic field created by any magnetic element inside the inner region - e. g. a permanent magnet -  should not leak to outside the region enclosed by the shell.
       
ii) The system formed by the enclosed region {\em plus the shell} should be magnetically undetectable from outside (no interaction -e. g. no magnetic force- with
any external magnetic sources).

In this work we will consider the case of a cylindrical cloak; results can be extended to other geometries. It was demonstrated in \cite{controlling,wood,yag}
that different sets of radially dependent values of radial and angular permeabilities, $\mu_\rho$ and $\mu_\theta$, yield a magnetic cloak behavior. An example is
\begin{equation}
\label{cylindrical-cloaking}
\mu_\rho={\rho-a\over{\rho}},
\mu_\theta={\rho\over{\rho-a}},
\end{equation}
represented in Fig. 1a, with the resulting field profile shown in Fig. 1b. Materials with such fine-tuned values of anisotropic permeabilities do not exist. It is easily seen from Eqs. (\ref{cylindrical-cloaking}) that $\mu_\theta=\infty$ and at the same time $\mu_\rho=0$ at the inner layer ($\rho=a$). This makes any practical implementation very difficult \cite{controlling}. In particular, condition (i) could not be fulfilled  because the field from any internal source would leak to the exterior. Therefore a strong (or an exact) validation of condition (i) for our antimagnet requires a radically new approach. This is what we present in this work. Two important new ideas would be required to achieve the design of a practical antimagnet fulfilling conditions (i) and (ii): firstly, the introduction of a new scheme for cloaking, requiring homogeneous (although anisotropic) parameters, which would involve a new transformation of space, and, secondly, the addition of an inner superconducting layer, which in the static case considered here would ensure $\mu=0$. In a final step we will present a general way of implementing the antimagnet design in practical cases.

\section{A cloaking design with homogeneous parameters}

As a first step, we want to study whether the values of $\mu$ derived for the magnetic cloak, which are impractical because they involve a fine-tuned continuous variation of anisotropic permeabilities, can be modified into a simplified scheme. 
Can a cloaking behaviour (in particular no field distortion in an exterior region) be produced with simpler permeability arrangements -even with a homogeneous permeability value? In the following, we will demonstrate that a whole family of homogeneous magnetic systems exist with the property that they produce a null effect on an externally applied magnetic field. Each of these systems -we call them homogeneous cloaks - is a cylindrical shell composed of an homogeneous (i. e., properties are the same in all the material points) anisotropic material.

Consider one of such cylinders infinitely long along the $z$ axis with radius $R_2$ and a central coaxial hole of radius $R_1<R_2$. This cylinder is made of a magnetic material with homogeneous radial and angular relative permeabilities, $\mu_\rho$ and $\mu_\theta$ respectively. A uniform external field $H_a$ is applied along the $y$ direction. We want to find the analytical expression for the field ${\bf H}$ in all regions with the condition that the applied field is not modified by the presence of the cylinder, in the region outside it. 

Inside the material there are neither free charges nor free currents. Then, $\nabla\cdot{\bf B}=0$ and $\nabla\times{\bf H}=0$ and given that neither $\mu_\rho$ nor $\mu_\theta$ depend on position, we obtain 
\begin{eqnarray}
\label{eq.1r}
\mu_\rho \rho \frac{\partial H_\rho}{\partial \rho} + \mu_\rho H_\rho + \mu_\theta \frac{\partial H_\theta}{\partial \theta} = 0,\\
\label{eq.1t}
\rho \frac{\partial H_\theta}{\partial \rho} + H_\theta - \frac{\partial H_\rho}{\partial \theta} = 0.
\end{eqnarray}

The boundary conditions for these equations are set by considering that both {\bf B} and {\bf H} equal the applied field values at $\rho=R_2$ and that there is a uniform field inside the hole. Then, by imposing continuity of the normal component of ${\bf B}$ and the tangential component of ${\bf H}$, we get
\begin{eqnarray}
	H_\rho(R_2) &=& H_a \frac{1}{\mu_\rho} \sin\theta, \\	
	H_\theta(R_2) &=& H_a \cos\theta, 
%	\\
%	H_\rho(R_1) &=& \alpha H_a \frac{1}{\mu_\rho} \sin\theta, \\
%	H_\theta(R_1) &=& \alpha H_a  \cos\theta.
\end{eqnarray}
The solution for the magnetic field inside the ring ($R_1\leq \rho \leq R_2)$ is found to be
\begin{eqnarray}
\label{eq.Hr}
	H_\rho(\rho,\theta)=H_a \mu_\theta \left(\frac{\rho}{R_2}\right)^{-1+\mu_\theta} \sin\theta, \\
\label{eq.Ht}
	H_\theta(\rho,\theta)=H_a  \left(\frac{\rho}{R_2}\right)^{-1+\mu_\theta} \cos\theta.
\end{eqnarray}
It is important to remark that this solution is valid only when the condition $\mu_\rho \mu_\theta = 1$ is fulfilled; otherwise, the problem has no solution. Therefore, we demonstrate that the condition of null external distortion of magnetic field is directly fulfilled as long as permeabilities $\mu_\rho$ and $\mu_\theta$ are the inverse of each other ($\mu_\rho\mu_\theta=1$). 

It is interesting to notice that the presented scheme for a homogeneous cloaking implicitly involves a space transformation different from that of Wood and Pendry \cite{wood}, which involved expanding a zero-dimensional point into a finite sphere corresponded (in our cylindrical case, expanding a central one-dimensional line into a cylindrical region, that from $\rho'=0$ to $\rho'=a$). This transformation basically  consists on, first, compressing the space from $\rho=R_0$ to $\rho=b$ into the space occupied by $\rho'=a$ and $\rho'=b$ using a radially symmetric function, and, second, expanding the space from $\rho=0$ to $\rho=R_0$ into the space from $\rho'=0$ to $\rho'=a$, using another radially symmetric function. $R_0$ is a positive constant ($R_0<a$). A detailed description and discussion of the transformation (which may also be used in the case of electromagnetic waves) will be presented elsewhere.

An example of such homogeneous cloak is shown in Fig. 1c, with the values  $\mu_\theta$=6 and $\mu_\rho$=1/6=0.1667 (solid lines in fig. 1a), and in Fig. 1d, with $\mu_\theta=10$ and  $\mu_\rho=$1/10 is shown; in both, the condition $\mu_\rho$=1/$\mu_\theta$ is enough to ensure no distortion of the magnetic field outside the shell. Interestingly, it can be demonstrated that the null distortion is also obtained for a non-uniformly applied magnetic field. As to the field in their interior, we see that increasing $\mu_\theta$ while maintaining the ratio $\mu_\rho\mu_\theta$=1 causes the magnetic field lines to be concentrated nearer the external surface of the shell. This means that such homogeneous cloaks do not have exactly zero magnetic field inside, although they do so approximately - the larger the value of $\mu_\theta$ the closer to zero the internal field.

In spite of achieving magnetic cloaking, these homogeneous cloaks are not antimagnets, because the magnetic field created by a source in its interior will leak to the exterior. To avoid this, we introduce the second key step in our idea: placing a superconducting layer at the inner surface of the cloak. Because $\mu=0$ for an ideal superconductor, it directly follows from the magnetostatic boundary conditions at the inner boundary of the superconductor that condition (i) is fulfilled. Introducing such a superconducting layer does not substantially modify the property of cloaking, as long as $\mu_\theta$ is sufficiently larger than 1, because in this case, as seen in Figs. 1c and 1d, magnetic field is excluded from the central part so a superconducting layer would not practically interact with the field created by external sources. In this way an antimagnet design is being outlined: an inner superconducting layer and an outer homogeneous shell. However, this scheme alone cannot yet solve our goal of a feasible antimagnet, because the material in the outer shell, even though it would have a constant permeability, would require fine-tuned anisotropic values (with $\mu_\rho=1/\mu_\theta$); such materials are not available.

\section{Antimagnet design and demonstration of its properties} 

Therefore, another important step is needed, and again help from metamaterials concepts will be used. Can we transform our homogeneous cloak with uniform and anisotropic parameters into one made with realistic materials?
We have attained this by modifying the homogeneous shell with constant anisotropic permeability into a discrete system of alternating layers of two different kinds: one type consisting simply of a uniform and isotropic ferromagnetic material with constant permeability ($\mu^{FM}_\rho=\mu^{FM}_\theta>1$) and a second type having a constant value of radial permeability $\mu^{SC}_\rho$ ($\mu^{SC}_\rho<1$) and $\mu^{SC}_\theta=1$. The first (isotropic) kind of layers could be a superparamagnet (i. e. ferromagnetic nanoparticles embedded in a non-magnetic media \cite{super}). The second kind could be realized with arrays of superconducting plates (the precise value tunable by changing distances between the plates) \cite{wood,magnus,ourAPL}; one of such arrays has actually been constructed and tested \cite{magnus}. Now we need to find the values of permeability for the two kinds of layers. To do this, we start with the values for a homogeneous cloak described above (for example, that in Fig. 1c) and then apply the following method: we select the value of the angular permeability ($\mu_\theta$=6) for the isotropic layers ($\mu^{FM}_\theta=\mu^{FM}_\rho=$6) and then reduce the value of $\mu^{SC}_\rho$ in the superconducting layers to a lower value than $1/\mu^{FM}_\rho$ (1/6) to compensate for the larger value of $\mu^{FM}_\rho$ in the isotropic layers. 

To demonstrate the validity of the method, we show in Fig. 2e the calculated response for a system with an inner superconducting layer surrounded by 10 outer alternating layers half of them with $\mu^{FM}_\theta=\mu^{FM}_\rho=6$ and the other half with $\mu^{SC}_\rho$=0.104 (and $\mu^{SC}_\theta=$1); the scheme cloaks a uniform static magnetic field with an impressive quality (as can be visualized in Fig. S1). This means that our goal of using only realistic material is fulfilled. It is only left to confirm that the resulting scheme acts indeed as an antimagnet and it does so for {\em any} applied field. This is demonstrated on the panels in Fig. 2, where we show the magnetic field of a single small magnet (basically a dipole field), the field of two such magnets (now the field has changed very much because of the interaction), and, finally, how surrounding one of the magnets with the antimagnet makes the field outside it unaffected - i. e. equal to the field of a single magnet. Besides the constant field and two dipole cases, we also show in Fig. 2d another example corresponding to the field created by a current line. In all cases the antimagnet is the same. We thus confirm that the antimagnet performs as a such for any applied field configuration. 
We also show in Fig. 2f that even when there is missing portion of the antimagnet, a reasonable shielding inside with a small field modification outside can be obtained. 

\section{Discussion}

The presented solution with 10 layers and $\mu^{FM}_\theta=6$ is not unique; starting from a different $\mu^{FM}_\theta=\mu^{FM}_\rho$ value (10, for example, as in Fig. 1d) and even with a different number of layers one can follow the described method and obtain similarly good antimagnet properties. Actually, 
although a good behavior is obtained for 10 layers as shown above, increasing the number of layers may be convenient when there is an applied field very spatially inhomogeneous or when we require a certain practical tolerance in the values of the permeabilities (more on this in Figs. 3 and S2).

We would like to make some remarks with practical consequences. First, we have considered that the superconductor is characterized by $\mu=0$, which is a good approximation for superconductors in the Meissner state. This would limit in principle the applicability of the antimagnet to applied fields less than the (thermodynamic or lower) critical field of the superconductor. However, a type-II superconductor (like most high-temperature ones) with a high critical-current density can produce a response very similar to a Meissner response (with currents circulating mainly in a thin layer at the surface) up to much larger fields \cite{araujo}.

Finally, constructing the described antimagnet with 10 layers may be feasible but difficult in practice. However, the same strategy can be applied to a much more simple design, by substituting all superconducting layers (except the central one) with layers with $\mu=1$ (the permeability of air or any non-magnetic material such as plastic). This change requires tuning the value of $\mu$ in the magnetic layers to a different value. An example of that is shown in Fig 4.  This simplified scheme works equally well than the described antimagnet when there is a uniform magnetic field, whereas its response gets worse with increasing applied field inhomogeneity. Moreover, in the latter case, increasing the number of layers does not bring a significant improvement of the antimagnet properties. Therefore, this simplified scheme with only a central superconducting layer and layers with homogeneous $\mu$ alternated with air may work well for applications in which the applied field has a small spatial variation.

\section{Conclusions}
In summary, we have presented a method to design hybrid superconductor-metamaterial devices that prevent any magnetic interaction with its interior while keeping the external magnetic field unaffected. Two important key ideas have been needed for achieving our goal: the design of a simplified cloak with homgeneous parameters, corresponding to a new space transformation, and the placement of a superconducting layer at the inner surface. Such an antimagnet would be passive and, provided that the superconductors are in the Meissner state \cite{ourAPL} and that the isotropic layers have a negligible coercivity (as if using superparamagnetic materials), also lossless. The strategy for antimagnet design presented in this work can be adapted to other geometries (e.g. spheres) or even to other forms of manipulating magnetic fields, such as magnetic field concentrators \cite{concentrator,yag}. Antimagnet devices may bring important advantages in fields like reducing the magnetic signature of vessels or in allowing patients with pacemakers or cochlear implants to be allowed to use medical equipment based on magnetic fields, such as magnetic resonance imaging MRI \cite{mri} or transcraneal magnetic stimulation \cite{trans}.  Moreover, by tuning one parameter like the working temperature of the device - below or above the critical temperature of the superconductor, for example - one could 'switch off and on' magnetism in a certain region or material at will, opening up room for some novel applications.

\section*{Methods}

The simulations have been performed with Comsol Multiphysics software, using the electromagnetics module (magnetostatics). All the presented results correspond to an infinitely long cylinder (with translational symmetry) with outer radius equal twice the inner radius ($b=2a$). We have checked that other dimensions yield similar results. Except when otherwise indicated, we have considered 10 outer layers plus an inner superconducting layer, all with the same thickness. The outest layer is of the magnetically isotropic type. The superconductor has been simulated assuming $\mu=0$. All permeabilities are understood to be relative permeabilities.
The pink region in Figs. 3 and 4 denoting the region in which the magnetic induction {\bf B} differs from the externally applied magnetic induction ${\bf B}_{\rm ext}$ is calculated as the points in space following the condition
\begin{equation}
\sqrt{{(B_{\rm x}-B_{\rm x,ext})^2+(B_{\rm y}-B_{\rm y,ext})^2}\over{(B_{\rm x,ext})^2+(B_{\rm y,ext})^2}}>0.01
\end{equation}

\section*{Acknowledgements}

We thank Spanish Consolider Project NANOSELECT (CSD2007-00041) for financial support.

%%%%%%%%%%%%%%%%%%%%%%%%%%%%%%%%%%%%%%%%%%%%%%%%%%%%%%%%%%

\newpage
\begin{figure}[htbp]
	\centering
		\includegraphics[width=1\textwidth]{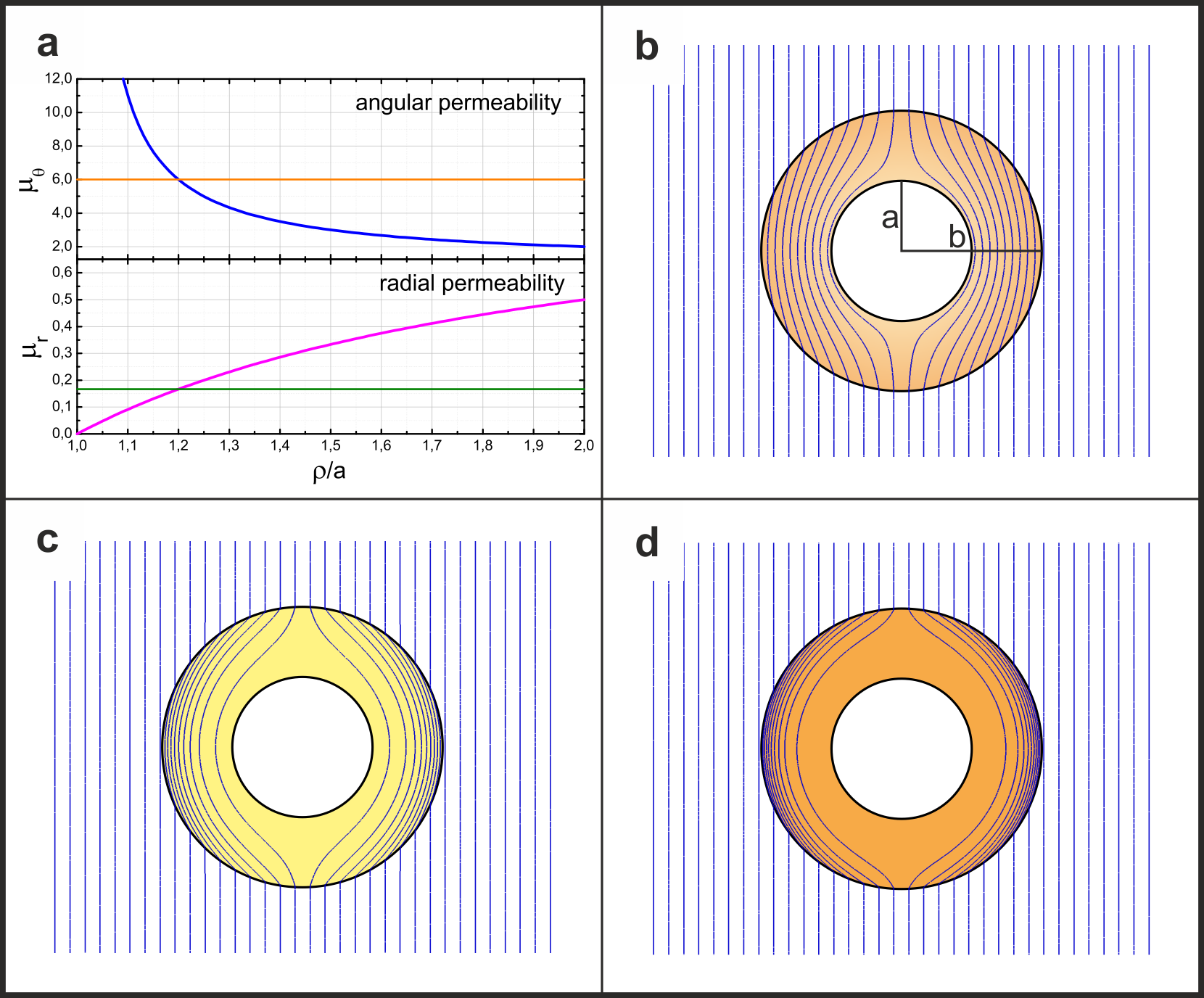}
\caption{{\bf Magnetic permeabilities and behavior of exact and approximate magnetic cloaks}. In panel ({\bf a}) radial dependence of radial and angular permeability for a cylinder of inner radius $a$ and outer radius $b=2a$; curves are the values for the exact cloak [Eq. (1)] and straight lines are the approximate constant values in panel ({\bf c}). In the rest of the panels, magnetic field lines for the ({\bf b}) exact cloak, ({\bf c}) approximate cloak with constant values $\mu_\theta$=6 and $\mu_\rho$=1/6=0.1667, and ({\bf d}) approximate cloak with constant values $\mu_\theta$=10 and $\mu_\rho$=1/10.}
	\label{fig1}
\end{figure}

\newpage
\begin{figure}[htbp]
	\centering
		\includegraphics[width=1\textwidth]{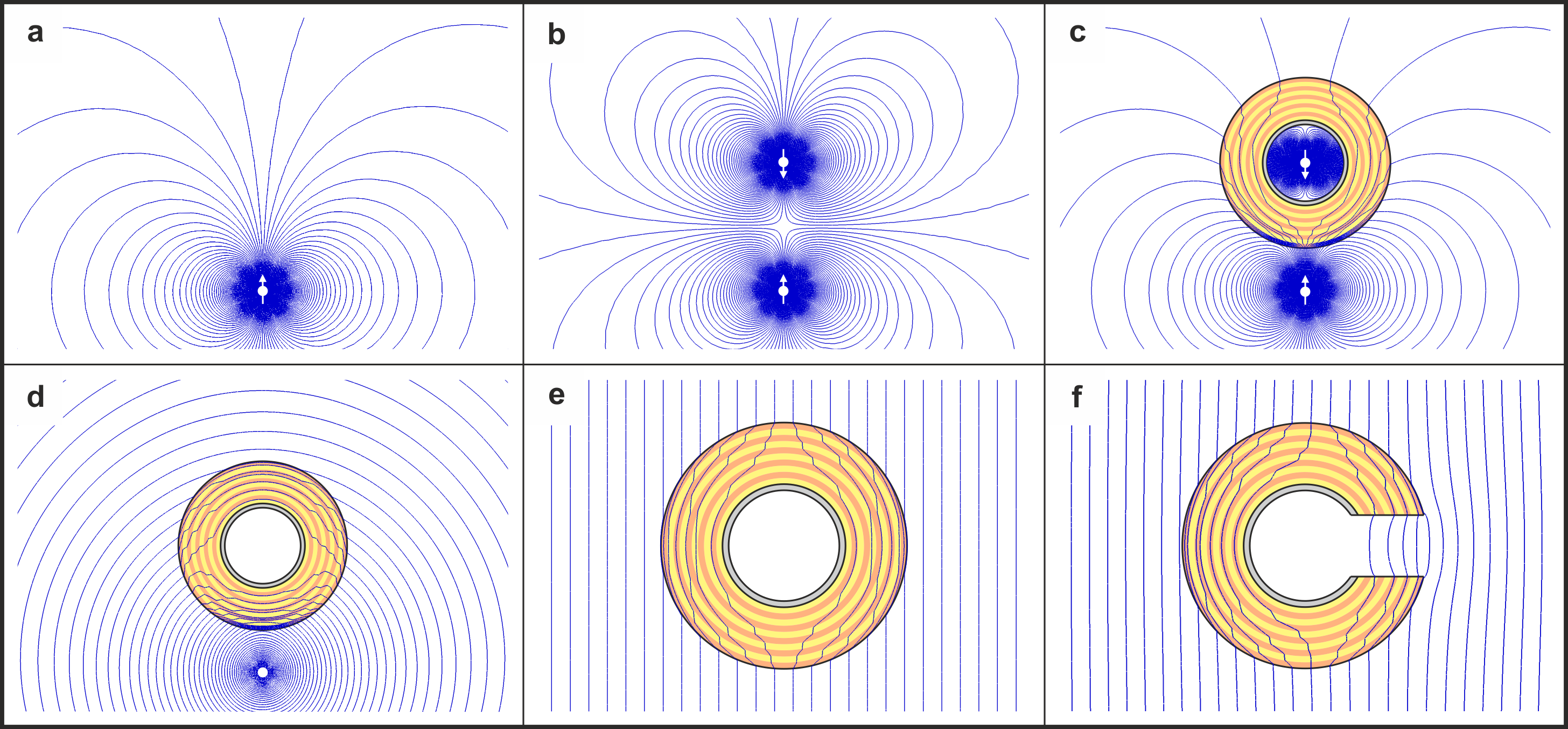}
	\caption{{\bf Display of the antimagnet behavior.} The magnetic properties of an antimagnet are visualized as follows. First, we show in panel ({\bf a}) the magnetic field lines for one uniformly magnetized cylindrical magnet. When a second magnet is added ({\bf b}) the magnetic field is distorted owing to the magnetic interaction between both magnets. When one of the two magnets is covered by the antimagnet ({\bf c}), then the magnetic field outside the region enclosed by the antimagnet is the same as that for a single magnet (as in panel a), demonstrating the two antimagnet properties: the field of the inner magnet does not leak outside the antimagnet shell and the field external to the antimagnet remains unperturbed independently of what it is contained in its interior. In panel ({\bf d}) and ({\bf e}) the same antimagnet is shown to behave as such in the field of a current carrying wire and a uniformly applied magnetic field, respectively. Panel ({\bf f}) shows that a rather good antimagnet behavior is maintained even when the shell is not closed. The antimagnet is composed of an inner superconducting layer ($\mu=0$) and 10 alternating outer layers of two kinds: one with $\mu^{FM}_\theta=\mu^{FM}_\rho=6$ and the other with $\mu^{SC}_\rho$=0.104.}
	\label{fig2}
\end{figure}
\newpage

\begin{figure}[htbp]
	\centering
		\includegraphics[width=1\textwidth]{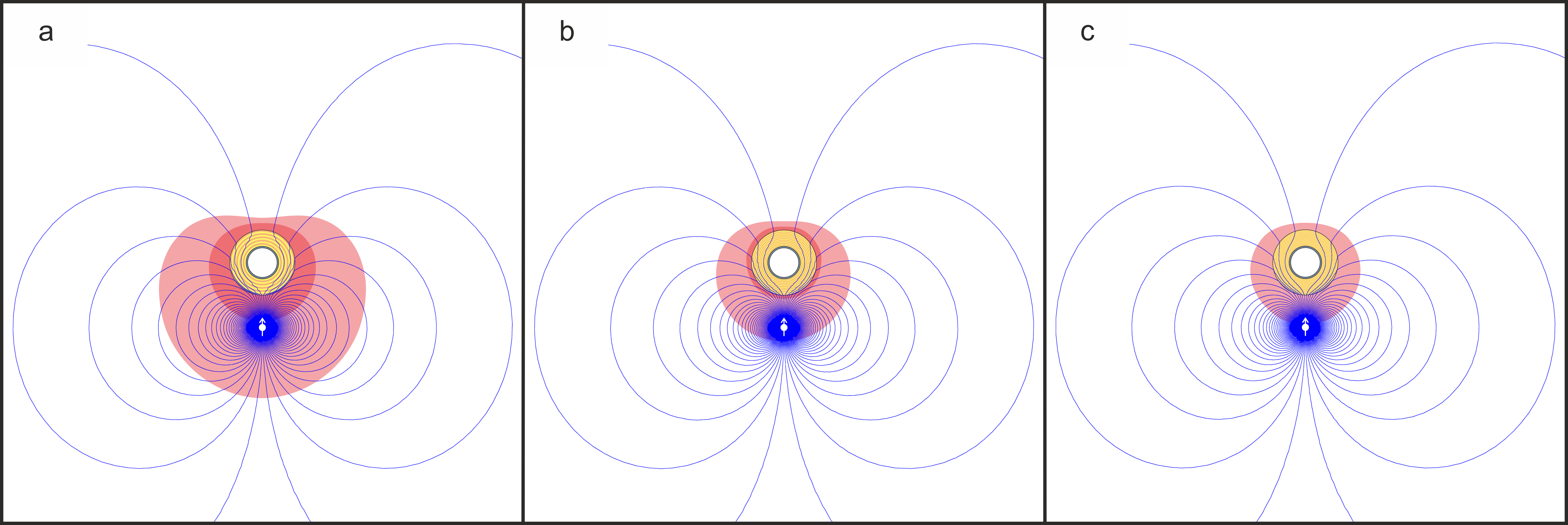}
	\caption{{\bf Optimizing the number of layers for the antimagnet.}
Response of antimagnets with -from left to right- 10, 20, and 30 layers to the field created by a near uniformly magnetized cylindrical magnet. The permeabilities of the magnetic layers are in all cases
$\mu^{FM}_\theta=\mu^{FM}_\rho=6$ whereas the superconducting ones are $\mu^{SC}_\rho$=0.104, 0.128, and 0.136, for the 10, 20, and 30 layer-cases, respectively. The light pink region indicates the zones for which the difference between the total field and the dipole field exceeds 1\% and the darker pink region when they exceed 3\%. }
\end{figure}

\newpage
\begin{figure}[htbp]
	\centering
		\includegraphics[width=1\textwidth]{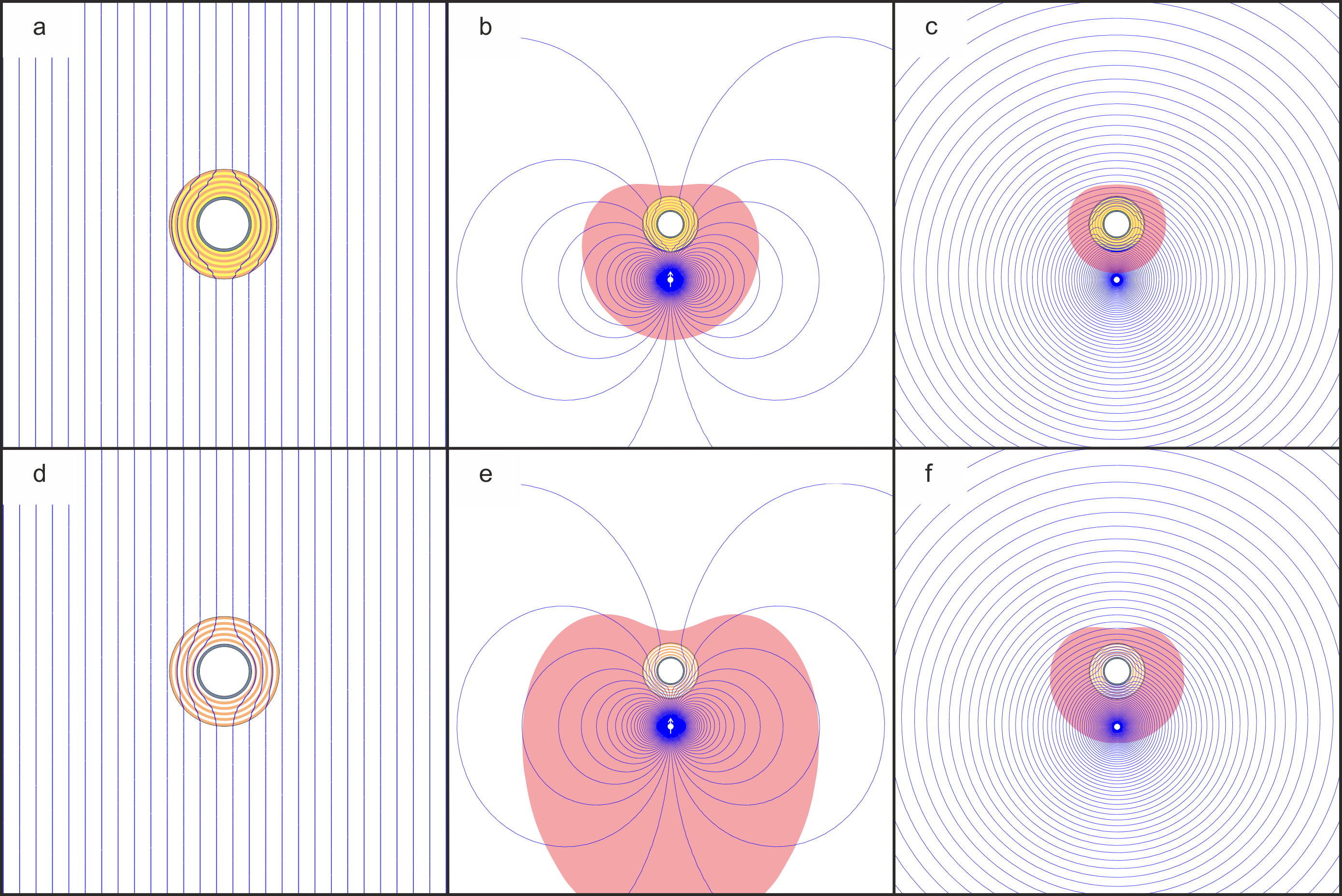}
	\caption{{\bf Simplified antimagnet with air ($\mu=1$) layers.}
Response of the antimagnet of 10 layers (as in Fig. 1) for a uniform applied field, a dipole-like field and the field of a current-carrying wire (top row, form left to right). For comparison, the same results are plotted in the bottom row for the simplified antimagnet case in which the permeability $\mu^{SC}_\rho$ of the superconducting layers - except the central one, which has $\mu=0$- has been set as 1 (and $\mu^{FM}_\theta=\mu^{FM}_\rho$=2.405 in this case). The pink regions indicate the zones for which the difference between the total field and the externally applied field exceeds 1\%. Both antimagnet and simplified designs work well for uniform applied field, but the region of distortion of the external field is large in the latter. }
\end{figure}

\end{document}